\documentclass[12pt,letterpaper]{article}
\pdfoutput=1

\usepackage{color,datetime,graphicx}
\usepackage{amsmath,amssymb,array,calc,rotating,epsfig,psfrag, amscd,cite}
\usepackage[colorlinks=false,
    linktoc=all, 
    pdfstartview=FitV,
    bookmarksopen=true]{hyperref}
\usepackage[english]{babel}

\usepackage[total={6.5in,8.75in},
top=1.1in,bottom=1.1in,left=1.2in,right=1.2in,includefoot]{geometry}

\newcommand{\rcite}{\cite}

\definecolor{darkblue}{rgb}{0, 0, 0.7}

\def\nfive{{n_5}}

\numberwithin{equation}{section}

\def\nn{\nonumber}

\begin{document}

\thispagestyle{empty}
 \begin{flushright}
 \end{flushright}
\vspace{0.5cm}
\begin{center}
  {\LARGE \bf{A note on integrability loss \\[8pt]in fuzzball geometries}}
 \vskip1.5cm 
\renewcommand{\thefootnote}{\fnsymbol{footnote}}
Maxim Emelin,  Stefano Massai
\vskip0.5cm
\textit{Dipartimento di Fisica e Astronomia ``Galileo Galilei''\\
  Universit\`a di Padova, Via Marzolo 8, 35131 Padova, Italy}\\
\vspace{0.25cm}
\&\\
\vspace{0.25cm}
\textit{INFN, Sezione di Padova\\
  Via Marzolo 8, 35131 Padova, Italy}
\vspace{0.5cm}
\vskip1cm
\end{center}

\begin{abstract}

\noindent
We study the dynamics of certain string configurations in a class of fivebrane supertube backgrounds. In the decoupling limit of the fivebranes, these solutions are known to admit an exact description in worldsheet string theory and  string propagation is integrable. For the asymptotically flat solutions, we prove, by using analytic tools of classical Hamiltonian systems, the non-integrability of classical string motion. This suggests that string dynamics in circular supertube geometries exhibit a regime of chaotic behaviour.
\end{abstract}

\clearpage


\section{Introduction}

In recent years there has been considerable progress in
constructing black hole microstate geometries (see \rcite{Bena:2022rna} for a
recent review) and in studying their dynamical
properties. In some special cases, it has been shown that the dynamics
of strings on such geometries is described by exact null-gauged WZW
models \rcite{Martinec:2017ztd,Martinec:2018nco,Bufalini:2021ndn}, which allow to study stringy aspects that are invisible in
supergravity. The simplest example is the circular NS5-F1
supertube. The exact description is found after taking the decoupling limit
for the  fivebranes. The resulting geometry interpolates between a
linear dilaton region in the UV and an orbifold of $AdS_3$ in the IR. The
null-gauging construction that ``adds back the one'' in the F1
harmonic function has been generalized to encompass a larger class of
Lunin-Mathur solutions in \rcite{Martinec:2020gkv}. It is natural to
ask to what extent one can extend these constructions in order to
describe the full backgrounds by similarly ``adding back the one'' in the fivebrane
harmonic function, effectively recoupling the flat space
asymptotics.

In this note we will show that there is a considerable difference
between the interpolation from the $AdS_3$ region to the linear dilaton region, and the
interpolation to the asymptotically flat region. By studying the
classical Hamiltonian system that describes the dynamics of strings on
such geometries, we will show that properties of certain string configurations in the
solutions with linear dilaton
asymptotics are consistent with integrability, in agreement with
the fact that they are described by an explicit gauged WZW model. On
the other hand, coupling back the asymptotically flat region has the
consequence of breaking this integrability.  We will prove this fact by using analytic tools
that have been developed in the context of Hamiltonian systems and
that have been already applied to study the (non)integrability of
strings in different contexts (see for example
\rcite{Basu:2011fw,Stepanchuk:2012xi,Roychowdhury:2017vdo,Nunez:2018qcj,Filippas:2019ihy,Roychowdhury:2020zer,Rigatos:2020igd}).

The notion of integrability of a classical Hamiltonian system, in the
sense of Liouville, relies on the behaviour of trajectories in phase space, as specified by
the Arnold-Liouville theorem. In particular, integrable systems are
non-chaotic. The transition to chaotic behaviour for small
perturbations around an integrable system is described by KAM
theory. The work of Ziglin provides effective criteria for the
integrability and non-integrability of Hamiltonian systems in terms of analytic
properties of the corresponding differential equations, through the analysis of the
monodromy group of the linearized system normal to an integrable plane of solutions.
These criteria have been simplified by the use of differential
Galois theory in what is known as the Ziglin-Morales-Ramis theory. The
main result is that a necessary condition for a system to be
integrable is that the Galois group of the normal variational equation
(NVE) has an abelian identity component. 
In practice, this means that one can prove non-integrability of a system by perturbing the equations of motion around a particular solution to linear order (while respecting the necessary constraints) and showing that the resulting equations do not admit Liouvillian solutions.

This criterion is
particularly useful because the work of Kovacic \cite{KOVACIC19863} provides an explicit algorithm, based
on  Picard-Vessiot theory, that allows to check the properties of the
Galois group of a given differential equation and in particular, either find Liouvillian solutions or prove their non-existence. This algorithm is
known as the Kovacic algorithm (see Appendix \ref{sec:K} for a brief
review). For a general introduction see for example
\rcite{MoralesBook}. We will apply these tools in the string theory context to check whether the dynamics of classical strings in supertube backgrounds are integrable or not.

The paper is organized as follows. In Section
\ref{sec:sugra} we describe the supergravity solutions for
supertubes and their decoupling limits. In Section
\ref{sec:stringmotion} we derive the equations of motion for a
particular configuration of strings moving in such geometries and we
derive the corresponding NVEs. We then
apply the Kovacic algorithm to these equations in order to check
the necessary condition for integrability of the system and find that it fails for the asymptotically flat backgrounds, implying non-integrability. In
Section \ref{sec:conclusions} we conclude and discuss some consequences of
our result. We relegate a review of the Kovacic algorithm and details of its application to our NVEs to Appendix \ref{sec:K}.




\section{Supergravity solutions}\label{sec:sugra}

In this Section we review the solutions corresponding to a circular
distribution of NS5 branes and to the circular fuzzball geometry in
the NS5 frame.

\subsection{Coulomb branch of NS5 branes}

The solution sourced by a collection of $n_5$ NS5 branes parallel to the
$0-5$ directions and placed at points
$x_n^{i}$, $i=6,7,8,9$ in the transverse space is:
\begin{align}
ds^2 &= \eta_{\mu \nu}dx^{\mu}dx^{\nu} + H dx^i dx_i \, , \\[8pt]
       e^{2\Phi} &= g_s^2 H \, , \nonumber\\[8pt]
  H_{ijk} &= -\epsilon^{l}_{ijk}\partial_lH \, , \nonumber
\end{align}
where the harmonic function $H$ is given by\footnote{We set $\alpha'=1$.}
\begin{equation}
  H =
  1+\sum_{n=1}^{n_5} \frac{1}{|x^i - x^i_n|^2} \, .
  \end{equation}
We consider the NS5 branes equally distributed on a circle of 
radius $a$ in the $(x_6, x_7)$
plane. It is useful to use the following coordinates on the transverse
$\mathbb{R}^4$:
\begin{align}
x_6 + x_7 = r e^{i \phi} \, ,\quad x_8 + i x_9 =
 R e^{i \psi} \, .
\end{align}
The fivebrane harmonic function can then be written as \rcite{Sfetsos:1998xd,Israel:2005fn}
\begin{equation}
H = 1+\frac{n_5\Lambda_{n_5}}{2 a r \sinh (\chi)} \, , 
\quad \Lambda_{n_5} = \frac{\sinh(n_5 \chi)}{\cosh(n_5 \chi) -
  \cos(n_5\phi)} \, ,
 \end{equation}
 with
 \begin{equation}
2 a r \cosh \chi = r^2 + R^2 +a^2 \, .
\end{equation}
We will consider the limit $\Lambda_{n_5} \rightarrow 1$, which corresponds
to a smeared distribution of branes along the circle.
Setting, in the notation of \rcite{Martinec:2017ztd}
\begin{align}
r = a \cosh \rho \sin \theta \, ,\quad R=a
  \sinh \rho \cos \theta \, ,
\end{align}
one finds
\begin{equation}
H = 1+\frac{n_5}{a^2\Sigma} \, ,\quad \Sigma = \cosh^2 \rho -
\sin^2\theta  \, ,
\end{equation}
and the transverse metric is:
\begin{align}
\label{NS5af}
ds^2_{\perp} &= 
(a^2\Sigma+n_5)\Bigl[d\rho^2 + d\theta^2 + \frac{1}{\Sigma}\Bigl(
{\cosh}^2\!\rho\sin^2\!\theta \,d\phi^2 +
               {\sinh}^2\!\rho\cos^2\!\theta \,d\psi^2\Bigr)\Bigr] \,
               .
\end{align}
In the decoupling limit of the fivebranes the metric and B-field  are:
\begin{align}
\label{NS5 coulomb}
ds^2_{\perp} &= 
n_5\Bigl[d\rho^2 + d\theta^2 + \frac{1}{\Sigma}\Bigl(
{\cosh}^2\!\rho\sin^2\!\theta \,d\phi^2 + {\sinh}^2\!\rho\cos^2\!\theta \,d\psi^2\Bigr)\Bigr],
\\[8pt]
B   &= \frac{n_5 \cos^2\theta \cosh^2\rho}{\Sigma} \, d\phi \wedge d\psi
\,.\qquad
%
\nonumber
\end{align}
String theory in this decoupled background is exactly solvable and it corresponds to the coset orbifold \cite{Sfetsos:1998xd,Giveon:1999tq}
\begin{equation}
\left( \frac{SL(2)}{U(1)}\times \frac{SU(2)}{U(1)} \right) \big/ Z_{n_5} \, ,
    \end{equation}
    or, equivalently, to the null gauging \cite{Israel:2004ir}
    \begin{equation}\label{ns5nullgauging}
\frac{SL(2)\times SU(2)}{U(1)_L\times U(1)_R} \, .
        \end{equation}

\subsection{NS5-P supertube}

Starting from the fivebrane configuration described in the previous
section one can construct a round supertube in the following way. We
compactify the $1-5$ directions on $T^4\times S^1$; call the $S^1$ direction $y$; we identify $y \sim y+ 2\pi R$. We can then tilt the NS5 brane strands along the $y,\phi$ directions and join them together to make
a helical profile. This system can be stabilised by spinning the fivebranes. The supergravity solution sourced by such a configuration
can be found by following a series of dualities starting from the F1-P
system \rcite{Lunin:2001fv} (see also \cite{Maldacena:2000dr}).  The result is 
\begin{align}
\label{NSPfull}
ds^2 &=  -du\, dv + ds_{\scriptscriptstyle\mathbf T^4}^2 
+ (a^2\Sigma+ \nfive)\Bigl(d\rho^2+d\theta^2 \Bigr)\nn\\[.1cm]
& \hskip0.6cm+  \frac{(a^2\Sigma+ \nfive)}{\Sigma} \Bigl( {\cosh}^2\!\rho\sin^2\!\theta \,d\phi^2 + {\sinh}^2\!\rho\cos^2\!\theta \,d\psi^2 \Bigr)\nn\\[.1cm]
& \hskip0.6cm
+\frac{1}{\Sigma} \Bigl[ \frac{2 k }{ R} \sin^2\!\theta \,dv\, d\phi + \frac{k^2}{\nfive R^2} dv^2 \Bigr],
\\[8pt]
B  
&= \frac{n_5 \cos^2\theta \cosh^2\rho}{\Sigma} d\phi \wedge d\psi +
\frac{ k  \cos^2\theta}{R\,\Sigma}  dv \wedge d\psi
~,\qquad~~~
e^{2\Phi}  = 1+ \frac{n_5}{a^2 \Sigma} \;.
\nn
\end{align}
where $u=t+y$, $v=t-y$ and the supertube radius is fixed in terms of the fivebrane and momentum charge by
\begin{equation}
a= \frac{\sqrt{Q_p Q_5} R}{k} \, .
    \end{equation}
The fivebrane decoupling limit of this solution admits an exact
description in worldsheet string theory as a null gauging that modifies \eqref{ns5nullgauging}\rcite{Martinec:2017ztd}
\begin{equation}\label{nullcoset}
\frac{SL(2)\times SU(2) \times S_y\times \mathbb{R}_t}{U(1)_L\times U(1)_R}
\, ,
\end{equation}
where we gauge two independent left and right null currents in the
upstairs group. After integrating out the gauge fields, one finds a
sigma model that describes the following metric and B-field:
\begin{align}
\label{smearedNS5Pmetric}
ds^2 &= \Bigl( -du\, dv + ds_{\scriptscriptstyle\mathbf T^4}^2 \Bigr)
+ \nfive\Bigl[d\rho^2+d\theta^2 +  \frac{1}{\Sigma} \Bigl( {\cosh}^2\!\rho\sin^2\!\theta \,d\phi^2 + {\sinh}^2\!\rho\cos^2\!\theta \,d\psi^2 \Bigr)\Bigr]
\nn\\[.1cm]
& \hskip 2cm 
+\frac{1}{\Sigma} \Bigl[ \frac{2 k }{ R} \sin^2\!\theta \,dv\, d\phi + \frac{k^2}{\nfive R^2} dv^2 \Bigr],
\\[8pt]
B  
&= \frac{n_5 \cos^2\theta \cosh^2\rho}{\Sigma} d\phi \wedge d\psi +
\frac{ k  \cos^2\theta}{R\,\Sigma}  dv \wedge d\psi
~.\qquad~~~
\nn
\end{align}
This indeed coincides with the fivebrane decoupling limit of the
solution \eqref{NSPfull}. 

\subsection{NS5-F1 supertube}

Starting from the metric of the NS5-P supertube, a T-duality along the $y$ circle brings the metric to the following
form:
\begin{align}
\label{NS5F1metric}
ds^2 &= ds_{\scriptscriptstyle\mathbf T^4}^2  -\Bigl( 1-\frac{\alpha^2}{\nfive\Sigma_1} \Bigr) du\:\! dv
+ (a^2 \Sigma +n_5)\Bigl( d\rho^2+ d\theta^2 \Bigr) \nonumber\\[.1cm]
& \hskip .6cm +\Bigl(a^2 \cosh^2\rho+n_5 + \frac{n_5 \sin^2\theta}{\Sigma_1}\Bigr) \sin^2\theta d\phi^2\nonumber\\[.1cm]
& \hskip .6cm+\Bigl(a^2 \sinh^2\rho+n_5 -\frac{n_5 \cos^2\theta}{\Sigma_1}\Bigr) \cos^2\theta d\psi^2
+ \frac{2\alpha}{\Sigma_1} \Bigl(  {\sin^2\!\theta \, dt\,d\phi + \cos^2\!\theta \, dy\,d\psi}  \Bigr)\, ,
\nonumber\\[8pt]
B  &= n_5 \cos^2\theta \Bigl(1+\frac{ \sin^2\theta }{\Sigma_1}\Bigr)   d\phi\wedge d\psi - \frac{\alpha^2}{n_5\Sigma_1} \, dt\wedge dy \nonumber\\
& \hskip .6cm 
{}+\frac{\alpha \cos^2\!\theta}{\Sigma_1}  dt\wedge d\psi
+\frac{\alpha  \sin^2\!\theta}{\Sigma_1}  dy\wedge d\phi~, \nonumber\\
& \hskip .6cm 
e^{2\Phi} = \frac{a^2\Sigma+n_5}{a^2\Sigma_1} ~,\quad 
\Sigma_1 = \frac{\alpha^2}{\nfive} + \Sigma \, ,
\end{align} 
where $\alpha = k \tilde R$, $a=\sqrt{Q_1 Q_5}/(k \tilde R)$, with $\tilde R$ the T-dual radius. The fivebrane decoupled limit of this solution is again described by
a null gauging of the form \eqref{nullcoset}. In fact T-duality just amounts to flipping a
sign in the left current being gauged. This decoupled geometry
provides an interpolation between a linear dilaton region in the UV to
an orbifold of $AdS_3$ in the IR. The fact that it arises as a gauged
WZW model implies, in particular, that string dynamics on this geometry is
integrable.

\section{String motion in NS5 backgrounds}\label{sec:stringmotion}

We now turn to the classical dynamics of strings in the
configuration of NS5 branes described in the previous section. In
order to illustrate the procedure for a simple case, we
first consider the NS5 decoupling limit of the circular distribution
of fivebranes, before proceeding to the asymptotically flat cases of the circular NS5 distribution and the supertube background.

\subsection{Decoupled limit}

We are interested in studying the motion of classical strings in the
geometry \eqref{NS5 coulomb}, which is obtained by taking a decoupling
limit of the asymptotically flat solution describing a circular
distribution of NS5 branes. The string action in conformal gauge is
\begin{equation}
\mathcal{L} = -\frac{1}{4\pi} \int d\tau d\sigma  \left[ \partial_a X^{\mu}\partial^a X^{\nu}G_{\mu
    \nu}  +\epsilon^{ab}\partial_a X^{\mu}\partial_b X^{\nu}
  B_{\mu\nu}\right] \, .
  \end{equation}
  The Virasoro constraints are
  \begin{align}
G_{\mu\nu}\partial_{\tau}X^{\mu} \partial_{\sigma}X^{\nu} = 0 \,  ,
    \qquad G_{\mu \nu} \left( \partial_{\tau} X^{\mu}
    \partial_{\tau}X^{\nu} + \partial_{\sigma} X^{\mu}
    \partial_{\sigma}X^{\nu} \right) = 0 \, .
  \end{align}
We consider a string configuration defined by the following embedding:
  \begin{equation}\label{embed}
t=t(\tau)  \, ,  \quad \rho=\rho(\tau) \, , \quad
\theta=\theta(\tau) \, \quad \psi=\psi(\sigma) \, ,\quad \phi =
\phi(\tau) \, \, .
\end{equation}
The string equations of motion for $t, \psi, \phi$ are solved by
\begin{equation}
t(\tau) = \tau \, ,\quad \psi(\sigma) = c_{\psi} \sigma \, , \quad
\dot{\phi}(\tau) = -c_{\psi}\tanh^2 \rho \, ,
\end{equation}
and the remaining system of equations for $\theta, \rho$ are
\begin{equation}
\ddot{\theta} = 0 \, , \quad \cosh^2\rho \ddot{\rho} +c_{\psi}^2
\tanh\rho = 0\, . 
  \end{equation}
This system admits the particular solution
\begin{equation}
\theta = 0\, , \quad  n_5 \dot{\rho}^2 = 
1 -c_{\psi}^2 n_5 \tanh^2\rho \, ,
  \end{equation}
where the last relation is found by solving the Virasoro
constraints. The NVE is derived by considering fluctuations around the
particular solution $\theta=0$. Setting $\theta= 0 + \delta \theta$
one trivially obtains the equation $\ddot{\delta \theta} =0$. Clearly
this equation admits the Liouvillian solution $\delta \theta = a \theta
+ b$, with $a, b$ arbitrary constants, which is consistent with
the fact that the decoupled geometry is described by a gauged WZW model and is therefore integrable. One could make different choices of the string embedding, and still obtain an integrable NVE equation. 
  
\subsection{Asymptotically flat circular NS5}

We now study string propagation in the full solution
\eqref{NS5af}. In what follows we set $a=1$ for convenience. We consider the same embedding \eqref{embed}, for which the equations of motion become:
\begin{align}
  0 & = 2 (n_5+\Sigma) \cos^2\theta \sinh^2\rho\psi'' \, ,\\[8pt]
  0 &= \psi' \Big(n_5  \dot{r}\sin^22\theta \sinh 2\rho 
 + n_5 \dot{\theta}\sinh^2 2\rho \sin 2\theta \Big) +
4\dot{\phi}\,\dot{r}\sin^2\theta \sinh 2\rho \Big[ \Sigma (n_5 + \Sigma) \nonumber\\[8pt]
  & -n_5 \cosh^2\rho\Big]+4\dot{\phi}\,\dot{\theta}\sin 2\theta \cosh^2\rho \Big[ \Sigma (n_5 + \Sigma)+n_5 \sin^2\theta\Big] + 4\ddot{\phi} \,\sin^2\theta \cosh^2\rho \Sigma
    (n_5+\Sigma)  \nonumber\\[8pt]
  0&= \psi'^2 \cos^2\theta  \Big[ \sinh2\rho \Sigma(n_5+\Sigma) -  2n_5 \cosh \rho\sinh^3\rho\ \Big]+ \dot{\rho}^2 \sinh 2\rho \Sigma^2 \nonumber\\[8pt]
     & - 2 \sin2\theta \Sigma^2 \dot{\rho}\dot{\theta}
     -\sinh2\rho \Sigma^2 \dot{\theta}^2
     -2n_5 \cos^2\theta \sinh 2\rho \Big[\cosh^2\rho - \Sigma\Big] \psi' \dot{\phi} \nonumber\\[8pt]
     & + \sin^2\theta \sinh 2\rho \Big[n_5 \cosh^2\rho -\Sigma (n_5+\Sigma)\Big]\dot{\phi}^2+ 2\Sigma^2 (n_5+\Sigma)\ddot{\rho}    \nonumber\\[8pt]
  0&= \psi'^2 \sin 2\theta  \sinh^2\rho \Big[ n_5 \cos^2\theta - \Sigma(n_5+\Sigma) \Big] + \dot{\rho}^2 \sin 2\theta \Sigma^2 \nonumber\\[8pt]
     & +2\sinh2\rho \Sigma^2 \dot{\rho}\dot{\theta}
     -\sin 2\theta \Sigma^2 \dot{\theta}^2
     +2n_5 \sin 2\theta \cosh^2\rho \Big[\cos^2\theta - \Sigma\Big] \psi' \dot{\phi} \nonumber\\[8pt]
     & - \cosh^2\rho \Big[2 n_5 \cos\theta \sin^3 \theta +\sin2\theta \Sigma (n_5+\Sigma)\Big]\dot{\phi}^2+ 2\Sigma^2 (n_5+\Sigma)\ddot{\theta}  \, . \nonumber
  \end{align}
  The Virasoro constraints give the condition
  \begin{align}
0 & = \psi'^2 (n_5+\Sigma) \cos^2\theta \sinh^2\rho +\dot{\rho}^2
    \Sigma(n_5+\Sigma)  \\[8pt]
    & - \dot{t}^2\Sigma + \dot{\theta}^2 \Sigma (n_5+\Sigma)
      +\dot{\phi}^2 (n_5+\Sigma)\cosh^2\rho \sin^2\theta \, . \nonumber
    \end{align}
This system has the following particular solution
\begin{align}
t&=\tau \, ,  \quad \psi = c_{\psi}\sigma\, , \nonumber \\
  \theta &=0 \, , \quad \phi = 0 \, , \\ 
  \dot{\rho}^2  &= \frac{1}{n_5 +\cosh^2 \rho} - c_{\psi}^2 \tanh^2\rho
 \, . \nonumber
\end{align}
As before, we perturb the equations of motion around $\theta=0$ in order to derive
the NVE. Setting $\theta(\tau)=0+\eta(\tau)$ we find the equation\footnote{Strictly speaking, the NVE is derived by expanding in all directions normal to the invariant submanifold of phase space to which our solution belongs. For our chosen solution, this requires also perturbing the $\rho$ and $\phi$ solutions while enforcing the Virasoro constraint. However, these perturbations can be shown to only contribute at quadratic order and thus do not enter the NVE.}
\begin{equation}
\mathcal{A}\partial_{\tau}^2 \eta + \mathcal{B}\partial_{\tau} \eta + \mathcal{C}\eta = 0 \, ,
\end{equation}
where
\begin{align}
  \mathcal{A} &=  n_5 +\cosh^2 \rho \\[8pt]
  \mathcal{B}& = 2 \cosh \rho \sinh\rho \left( \frac{1-c_\psi^2 \sinh^2 \rho - n_5 c_\psi^2 \tanh^2 \rho}{n_5 + \cosh^2 \rho} \right)^{1/2}\\[8pt]
  \mathcal{C} &= \frac{1-c_\psi^2 \sinh^2 \rho - n_5 c_\psi^2 \tanh^2 \rho}{n_5 + \cosh^2 \rho} - c_\psi^2 \sinh^2 \rho - n_5 c_\psi^2 \tanh^4 \rho  \, ,
\end{align}
where $\Sigma = n_5 + \Sigma$. By changing variable $\tau
\rightarrow x = \cosh \rho(\tau)$ we bring the equation in the form
\begin{equation}
\mathcal{A}\dot{x}^2 \partial_{x}^2  \eta + (\mathcal{A} \ddot{x} +\mathcal{B}\dot{x})\partial_{x} \eta + \mathcal{C} \eta = 0 \, .
  \end{equation}
  Finally, we can bring the NVE in normal form by defining $\xi = g
  \eta$, with
  \begin{equation}
b g - 2 g' = 0 \, .
    \end{equation}
    We then get
    \begin{equation}
\xi'' - U \xi = 0 \, ,
\end{equation}
with $U= F/G$,
\begin{align} \label{finGenEq}
  F&= -c_{\psi }^4 x^{16} + \left(-4 n_5 c_{\psi }^4-4 c_{\psi }^4-2 c_{\psi }^2\right) x^{14}  \\& +  \left(-10 n_5^2 c_{\psi }^4-16 n_5 c_{\psi }^4-2 n_5 c_{\psi }^2+14 c_{\psi }^4+14 c_{\psi }^2\right) x^{12} \nonumber \\& +  \left(-12 n_5^3 c_{\psi }^4-8 n_5^2 c_{\psi }^4+56 n_5 c_{\psi }^4+6 n_5^2 c_{\psi }^2+34 n_5 c_{\psi }^2-12 c_{\psi }^4-18 c_{\psi }^2+2 n_5-6\right) x^{10} \nonumber \\ 
  &+ \big(-5 n_5^4 c_{\psi }^4+16 n_5^3 c_{\psi }^4+60 n_5^2 c_{\psi }^4-48 n_5 c_{\psi }^4+6 n_5^3 c_{\psi }^2 \nonumber \\&\qquad +14 n_5^2 c_{\psi }^2-50 n_5 c_{\psi }^2+3 c_{\psi }^4+6 c_{\psi }^2-n_5^2-14 n_5+3\big) x^8 \nonumber \\ 
  &+  \left(12 n_5^4 c_{\psi }^4+8 n_5^3 c_{\psi }^4-56 n_5^2 c_{\psi }^4+12 n_5 c_{\psi }^4-6 n_5^3 c_{\psi }^2-34 n_5^2 c_{\psi }^2+18 n_5 c_{\psi }^2-2 n_5^2+6 n_5\right) x^6 \nonumber \\
  &+ \left(-10 n_5^4 c_{\psi }^4-16 n_5^3 c_{\psi }^4+14 n_5^2 c_{\psi }^4-2 n_5^3 c_{\psi }^2+14 n_5^2 c_{\psi }^2\right) x^4  \nonumber \\&+  \left(4 n_5^4 c_{\psi }^4+4 n_5^3 c_{\psi }^4+2 n_5^3 c_{\psi }^2\right) x^2 - n_5^4 c_{\psi }^4 \\[8pt]
G&= 4 x^2 \left(x^2-1\right)^2 \left(n_5+x^2\right){}^2 \left(n_5 c_{\psi }^2 \left(x^2-1\right) +x^2 \left( c_{\psi }^2\left(x^2-1\right)-1\right)\right){}^2 \nonumber
\end{align}
In this form one can implement Kovacic's algorithm (see Appendix
\ref{sec:K}) and see that the algorithm
fails, and thus the NVE does not admit, for generic values of the
parameters, a Liouvillan solution. This in turn implies that the
Hamiltonian system describing the dynamics of strings in the
background sourced by a circular array of NS5 branes is not
integrable.

\subsection{Supertube background}
We can repeat the process above for the NS5-P supertube solution. The equations of motion are more complicated due to the additional metric and B-field components. For the decoupled case, one can find a solution analogous to what we had in the decoupled circular NS5 case. We have checked that the corresponding NVE, despite being significantly more complicated, does in fact produce Liouvillian solutions, as expected.
For the asymptotically flat case, starting with the same initial string embedding ansatz \eqref{embed} we find a particular solution to the string equations of motion of the form
\begin{equation} \label{STeqn}
    \begin{aligned}
     t=&\tau \, , \quad \psi = c_{\psi}\sigma = \frac{k}{n_5 R} \sigma , \\ 
     \quad y &=0 \, ,\quad   \theta =0 \, , \quad \phi = 0 \, , \\ 
  \dot{\rho}^2(\tau) & = \frac{n_5 \left(n_5 R^2-k^2\right)-k^2 \sinh^2 \rho}{n_5^2 R^2 \left(n_5+\cosh^2 \rho\right)}
 \, .
    \end{aligned}
\end{equation}
Note that the previously unconstrained $c_{\psi}$ now acquires a fixed value that depends on the supertube  parameter $k/R$.
Expanding around this solution as $\theta(\tau) = 0+ \eta(\tau)$ we find, as before, the NVE of the form
\begin{equation}
    \mathcal{A}\partial_{\tau}^2 \eta + \mathcal{B}\partial_{\tau} \eta + \mathcal{C}\eta = 0 \, ,
\end{equation}
where now
\begin{equation}
    \begin{aligned}
         \mathcal{A} &= n_5+ a^2 \cosh^2 \rho\\[8pt]
  \mathcal{B}& =  a^2 \sinh(2\rho) \left(\frac{n_5(n_5 R^2 - k^2) - a^2 k^2 \sinh^2 \rho}{n_5^2 R^2 (n_5+ a^2 \cosh^2 \rho )} \right)^{1/2}\\[8pt]
  \mathcal{C} &= -\frac{a^4 k^2 \sinh ^2 \rho  (\cosh (2 \rho )+3)+4 a^2 k^2 n_5 \cosh ^2 \rho +2 n_5^2 \left(k^2-a^2 R^2\right)}{2 n_5^2 R^2 \left(a^2 \cosh ^2 \rho +n_5\right)} 
  \, ,
    \end{aligned}
\end{equation}
and $\rho$ is a function of $\tau$ obtained by solving the last equation in  \eqref{STeqn}. Through the same variable transformations as in the circularn NS5 case, we reduce the NVE to an equivalent equation in normal form:
\begin{equation}
    \xi''(x)- U(x)\xi(x) = 0
\end{equation}
with $U=F/G$ and \footnote{For the sake of notational simplicity we display $U$ by setting $a=1$.}
\begin{equation} \label{STNVE}
\begin{aligned}
    F&=-k^4 x^{10} +\left(-4 k^4 n_5-4 k^4-2 k^2 n_5^2 R^2\right)x^8  \\ 
    &+ \left(-10 k^4 n_5^2-14 k^4 n_5+14 k^4-2 k^2 n_5^3 R^2+14 k^2 n_5^2 R^2\right)x^6 \\
    &+\big(-12 k^4 n_5^3-10 k^4 n_5^2+42 k^4 n_5-12 k^4+6 k^2 n_5^4 R^2 \\
    &\qquad +40 k^2 n_5^3 R^2-18 k^2 n_5^2 R^2+2 n_5^5 R^4-6 n_5^4 R^4\big) x^4 \\
    & + \big(-5 k^4 n_5^4+2 k^4 n_5^3+29 k^4 n_5^2-30 k^4 n_5+3 k^4+6 k^2 n_5^5 R^2+20 k^2 n_5^4 R^2 \\
    &\qquad -44 k^2 n_5^3 R^2+6 k^2 n_5^2 R^2 - n_5^6 R^4-14 n_5^5 R^4+3 n_5^4 R^4\big)x^2  \\ 
    & -2 n_5^6 R^4+6 n_5^5 R^4 + 2 k^4 n_5^4+4 k^4 n_5^3-12 k^4 n_5^2+6 k^4 n_5-14 k^2 n_5^4 R^2+12 k^2 n_5^3 R^2 \\
    G&= 4 \left(x^2-1\right)^2 \left(n_5+x^2\right){}^2 \left(k^2 \left(n_5+x^2-1\right)-n_5^2 R^2\right){}^2 \, .
\end{aligned}
\end{equation}
Once again, the Kovacic algorithm fails to produce a solution. This proves the absence of Liouvillian solutions to the NVE and thus the non-integrability of string dynamics in the full NS5-P supertube background. Since integrability is preserved by T-duality, this result also implies the non-integrability of the full NS5-F1 supertube solution \eqref{NS5F1metric}.

\section{Conclusions}\label{sec:conclusions}

In this note we proved that the Hamiltonian system that
describes the dynamics of strings on the  asymptotically flat circular NS5-F1 fuzzball geometry is
not integrable, in contrast with the case of the fivebrane decoupling limit, where an explicit description in terms of a gauged WZW
model is available. 

This result implies that, although a gauged sigma model
description along the lines of \rcite{Martinec:2020gkv} 
that describes an asymptotically flat fuzzball might be found, the dynamics of
strings in this case is expected to be
considerably more intricate.
In particular, we expect that at least in some regimes it will exhibit
a chaotic behaviour. It would be interesting to show explicitly that the
dynamics of strings on the circular two-charge geometry is chaotic by
computing the typical signatures of chaos on the phase space of the
system by a numerical analysis of the equations of motion and the computation of the Lyapunov exponents (see for
example
\rcite{PandoZayas:2010xpn,Basu:2011dg,Basu:2011di,Nunez:2018qcj} for a
similar analysis in different contexts). In
particular, one could explore the idea of using the Lyapunov exponent
to distinguish fuzzballs from black hole solutions. Some aspects
of chaos on circular fuzzball geometries (in the S-dual D1-D5 frame) have been reported in
\rcite{Bianchi:2020des}, although their analysis was limited to
the point particle limit, i.e.  geodesic motion.

It could also be interesting to set up a perturbative expansion of the
worldsheet models for the asymptotically linear region
solutions that describes the coupling to flat
space, and to see if one can access some signatures of 
chaotic behaviour, and to explore the consequences for flat
space holography.


\section*{Acknowledgements}

The work of SM is supported by the MIUR program for young researchers ``Rita Levi Montalcini''.
\appendix




\section{Kovacic algorithm}\label{sec:K}

Consider a second order linear ODE in normal form
\begin{equation}
\xi''(x) - U(x) \xi(x) = 0 \, ,
\end{equation}
where $U(x)$ is a rational function. The Kovacic
algorithm provides an explicit way to produce Liouvillian solutions if
they exists, or prove that they do not exist, that can easily be implemented in a programming language. Here we review the algorithm itself. Proofs of all the steps can be found in the original paper \cite{KOVACIC19863}. 

In the following, the ``order at infinity'' of $U(x)$ will refer to its order as a zero, i.e. the difference between the highest power of $x$ in the denominator and the highest power of $x$ in the numerator, while the order of the pole $c$ is the highest power of $(x-c)^{-1}$ in the Laurent expansion around $c$. 

There are three cases to consider, corresponding to three types of possible solutions to the ODE. The necessary conditions for these three cases are as follows:
\begin{itemize}
\item Case 1: Every pole of $U$ has even order, or else has order 1. At infinity,
  the order of $U$ is even, or else is greater than 2.
  
\item Case 2: $U$ has at least one pole of odd order greater
  than 2, or else has order 2.

  \item Case 3: The order of poles of $U$ does not exceed 2, and the order of $U$ at infinity is at least 2. In this case the partial fraction decomposition of $U$ can be written as
  \begin{equation}
      U(x) = \sum_i \frac{\alpha_i}{(x-c_i)^2} + \sum_j \frac{\beta_j}{x-d_j}
  \end{equation}
  and we further require that  $\sum_j \beta_j = 0$, $\sqrt{1+4 \alpha_i} \in \mathbb{Q}$ $i$ and furthermore that
  \begin{equation}
      \sqrt{1+4\left(\sum_i \alpha_i + \sum_j \beta_j d_j \right)} \in \mathbb{Q}
  \end{equation}
  
\end{itemize}

  If none of the above are satisfied by $U$, then the equation does not admit Liouvillan solutions, and the system whose NVE is given by the above equation is not integrable. If the necessary conditions for any of the cases are satisfied, we must check the cases sequentially. In each case we construct several candidate algebraic equations based on the behaviour of $U$ at its poles and at infinity. The exact procedure varies by case, which is reviewed below. The solutions to these algebraic equations, if they exist, are then used to construct the solution to the original ODE. If the solution doesn't exist, we move to the next available case. If all cases fail, there are no Liouvillian solutions to the ODE.

The general equation \eqref{finGenEq} has order at infinity of $2$ and several poles each of order $2$. This means all three cases have their necessary conditions satisfied and must be checked sequentially.

For demonstration purposes, we choose $c_t = 1, c_\psi=1, n_5 = 1$ in \eqref{finGenEq} so that
\begin{equation} \label{Vdemo}
    U =-\frac{x^{16}+10 x^{14}-42 x^{10}+10 x^8+42 x^6-10 x^2+1}{4 \left(x^9-x^7-2 x^5+x^3+x\right)^2}
\end{equation}
but the final non-integrability result holds for general values. The nine order 2 poles are located at $0, \pm 1, \pm i$ and $\pm x_{\pm}$ where 
\begin{equation}
    x_{\pm} = \sqrt{\frac{1}{2} (\sqrt{5} \pm 1)}.
\end{equation}
For other values of $n_5$ the poles at $\pm i$ and $\pm x_\pm$ will move along the imaginary axis, but never become degenerate, so the orders of the poles do not change.
\subsection*{Case 1}
For each pole, denoted $c$, we compute three pieces of data $\{[\sqrt{U}]_c , \alpha_c^+ , \alpha_c^- \}$ in the following way
\begin{itemize}
    \item 
    If the order of the pole is 1, then 
    \begin{equation}
        \{[\sqrt{U}]_c , \alpha_c^+ , \alpha_c^- \} = \{ 0 , 1 ,1 \}
    \end{equation}
    \item 
    If the order of the pole is 2, and $b$ is the coefficient of $(x-c)^{-2}$ in the expansion around $c$, then 
    \begin{equation}
        \{[\sqrt{U}]_c , \alpha_c^+ , \alpha_c^- \} = \{ 0 \ , \ \frac12 + \frac12 \sqrt{1+4b} \ , \ \frac12 - \frac12 \sqrt{1+4b} \}
    \end{equation}
    \item If the order of the pole is $2\nu \geq 4$ then $[\sqrt{U}]_c$ is the polar part of the Laurent expansion of $\sqrt{U}$ around $c$. Let $a$ be the coefficient of $(x-c)^{-\nu}$ in $[\sqrt{U}]_c$ and $b$ be the difference of coefficients of $(x-c)^{-\nu-1}$ in the expansions of $U$ and $([\sqrt{U}]_c)^2$ respectively. Then
    \begin{equation}
        \alpha_c^\pm = \frac12 \left( \nu \pm \frac{b}{a} \right)
    \end{equation}
\end{itemize}
We also compute $ \{[\sqrt{U}]_\infty , \alpha_\infty^+ , \alpha_\infty^- \}$ in the following way
\begin{itemize}
    \item 
    If the order at infinity of $U$ is greater than 2, then
    \begin{equation}
        \{[\sqrt{U}]_\infty , \alpha_\infty^+ , \alpha_\infty^- \} = \{ 0 , 0 , 1 \}
    \end{equation}
    \item 
    If the order at infinity of $U$ is equal to 2, and $b$ is the coefficient of $x^{-2}$ in the expansion around infinity, then
    \begin{equation}
         \{[\sqrt{U}]_\infty , \alpha_\infty^+ , \alpha_\infty^- \} =  \{ 0 \ , \ \frac12 + \frac12 \sqrt{1+4b} \ , \ \frac12 - \frac12 \sqrt{1+4b} \}
    \end{equation}
    \item
    If the order at infinity of $U$ is $-2\nu \leq 0$ then $[\sqrt{U}]_\infty$ is the sum of non-positive powers of $x$ in the expansion around infinity. Let $a$ be the coefficient of $x^\nu$ in $[\sqrt{U}]_\infty$ and $b$ be the difference of coefficients of $x^{\nu-1}$ in the expansions of $U$ and $([\sqrt{U}]_\infty)^2$ respectively. Then
    \begin{equation}
        \alpha_\infty^\pm = \frac12 \left( -\nu \pm \frac{b}{a} \right) \ .
    \end{equation}
\end{itemize}
Having constructed the above data, we compute
\begin{equation} \label{deqnc1}
    d = \alpha_\infty^{\pm} - \sum_{\{c\}} \alpha_c^{\pm}
\end{equation}
for all choices of $\pm$ for each $c$ independently. If $d$ is a non-negative integer, then for the \emph{same choice of signs} we compute
\begin{equation} \label{omegaeqnc1}
    \omega = \pm [\sqrt{U}]_\infty \sum_{\{c\} } \pm [\sqrt{U}]_c + \frac{\alpha_c^{\pm}}{x-c} \ .
\end{equation}
For each such $\omega$ we search for a monic polynomial $P$ of degree $d$ that satisfies
\begin{equation} \label{Peqn}
    P^{\prime \prime} + 2 \omega P^\prime + (\omega^\prime + \omega^2  - U )P =0
\end{equation}
If such a polynomial exists, then $P e^{\int \omega}$ is a solution to the original ODE. If it doesn't exist for any of the $\omega$'s, then case 1 fails.

For our example we obtain for the poles
\begin{equation}
\begin{aligned}
    \{[\sqrt{U}]_0 , \alpha_0^+ , \alpha_0^- \} &= \{ 0 , \ \frac12 , \ \frac12 \} \\
   \{[\sqrt{U}]_c , \alpha_c^+ , \alpha_c^- \} &= \{ 0 , \ \frac34 , \ \frac14 \} \qquad   c=\pm1 , \pm i, \pm x_{\pm} 
\end{aligned}
\end{equation}
and at infinity we have
\begin{equation}
     \{[\sqrt{U}]_\infty , \alpha_\infty^+ , \alpha_\infty^- \} = \{ 0, \frac12, \frac12 \}
\end{equation}
It's easy to see that no choice of signs produces a non-negative $d$ 
 as defined in \eqref{deqnc1}, since the $\alpha_0^{\pm}$ already cancel positive contribution from the point at infinity. This rules out case 1.

\subsection*{Case 2}

For case 2, we define a set of integers $E_c$ associated to each pole of $U$, denoted $c$, as follows: 
\begin{itemize}
    \item 
    If the order of the pole is 1, then $E_c = {4}$
    \item
    If the order of the pole is 2 and $b$ is the coefficient of $(x-c)^{-2}$ in the partial fraction decomposition of $U$, then
    \begin{equation}
        E_c = \{ 2+k\sqrt{1+4b} \ | \ k \in \{ -2,0,2 \} \} \cap \mathbb{Z}
    \end{equation}
    \item 
    If the order of the pole is $\nu > 2$ then $E_c = \{ \nu \}$.
\end{itemize}
We also define $E_{\infty}$ as follows
\begin{itemize}
    \item If the order of $U$ at infinity is greater than 2, $E_\infty = \{ 0 , 2 , 4 \}$
    \item 
    If the order of $U$ at infinity is equal to 2 and $b$ is the coefficient of $x^{-2}$ in the expansion around infinity, then
    \begin{equation}
        E_{\infty} = \{ 2+k\sqrt{1+4b} \ | \ k \in \{ -2,0,2 \} \} \cap \mathbb{Z}
    \end{equation}
    \item
    If the order at infinity is $\nu < 2$ then $E_\infty = \{ \nu \}$
\end{itemize}
We now consider all families $\{ e_c \} \cup \{ e_\infty \} $ where $e_c \in E_c$ and $e_\infty \in E_\infty$. For each family where at least one of the integers is odd we compute
\begin{equation} \label{deqnc2}
    d = \frac12 \left( e_\infty - \sum_{\{c \} } e_{c} \right)
\end{equation}
For each family that gives a non-negative integer $d$ we define
\begin{equation}
    \theta = \frac12 \sum_{ \{c \} } \frac{e_c}{x-c}
\end{equation}
and search for a monic polynomial $P$ of degree $d$ that satisfies
\begin{align} \label{Peqnc2}
    P^{\prime \prime \prime} &+ 3 \theta P^{\prime \prime} + (3\theta^2 + 3 \theta^\prime - 4 U)P^\prime \\ 
    &+(\theta^{\prime \prime} + 3 \theta \theta^\prime + \theta^3 - 4 U \theta - 2 U^\prime)P = 0 \ .
\end{align}
If such a solution is found, define $\phi = \theta + P^\prime / P$ and let $\omega$ be a solution to
\begin{equation}
    \omega^2 + \phi \omega + (\frac12 \phi^\prime + \frac12 \phi^2 - U) = 0
\end{equation}
Then $e^{\int \omega}$ is a solution to the original ODE. If no polynomials $P$ can be found, then case 2 fails.

For our example, we have 
\begin{equation}
\begin{aligned}
     E_0 &= \{ 2\} \\
     E_c &= \{ 2, 3, 1 \} \qquad c=\pm 1, \pm i , \pm x_\pm \\
     E_{\infty} & = \{ 2 \}
\end{aligned}
\end{equation}
Again, no combination results in a positive $d$ as defined in \eqref{deqnc2}. This rules out case 2.
\subsection*{Case 3}

Choose $n = 4,6$ or $12$. In principle one can immediately start with $n=12$, but it is easier to consider the cases sequentially if one expects to find solutions. To rule out solutions we must verify the $n=12$ case.

Similar to case 2, define a set of integers $E_{c}$ for each pole $c$ as follows:
\begin{itemize}
    \item 
    If the order of the pole is 1, then $E_c = \{ 12 \} $
    \item 
    If the order of the pole is 2, and $b$ is the coefficient of $(x-c)^{-2}$ in the partial fraction decomposition of $U$, then
    \begin{equation}
        E_c = \{ 6 + \frac{12k}{n}\sqrt{1+4b} | k \in {-n/2, ... -1, 0 ,1, ... , n/2} \} \cap \mathbb{Z} 
    \end{equation}
\end{itemize}
Let $\gamma$ be the coefficient of $x^{-2}$ in the expansion of $U$ around infinity. Then define
\begin{equation}
        E_\infty = \{ 6 + \frac{12k}{n}\sqrt{1+4\gamma} | k \in {-n/2, ... -1, 0 ,1, ... , n/2} \} \cap \mathbb{Z} 
    \end{equation}
We now consider all families $\{ e_c \} \cup \{ e_\infty \} $ where $e_c \in E_c$ and $e_\infty \in E_\infty$. For each family we compute
\begin{equation} \label{deqnc3}
    d = \frac{n}{12} \left( e_\infty - \sum_{\{c \} } e_{c} \right) \ .
\end{equation}
For each family that gives a non-negative integer $d$ we define
\begin{equation}
    \theta = \frac{n}{12} \sum_{ \{c \} } \frac{e_c}{x-c}
\end{equation}
and we also define $S = \prod_{\{c \} } (x-c)$. Next, consider the recursive relation
\begin{equation}
    P_{i-1} = -S P_i^\prime + ((n-i) S^\prime - S \theta)P_i - (n-i)(i+1)S^2 U P_{i+1} \ .
\end{equation}
We must now look for a monic polynomial $P$ such that for $P_n = -P$, the $P_{-1}$ generated by the above relation is identically zero. If such a polynomial is found, then let $\omega$ be a solution to
\begin{equation}
    \sum_{i=0}^n \frac{S^i P_i}{(n-i)!}\omega^i = 0 \ .
\end{equation}
Then $e^{\int \omega}$ is a solution to the original ODE. If no such $P_n$ can be found (for all choices of $n$), case 3 fails.

In our example, taking directly $n=12$, we have
\begin{equation}
    \begin{aligned}
        E_0 &= \{ 6 \} \\
        E_c &= \{ 3,4,5,6,7,8,9 \} \qquad c= \pm 1, \pm i , \pm x_\pm \\
        E_\infty & = \{ 6 \}
    \end{aligned}
\end{equation}
And once again, no combination can give a non-negative $d$ as defined in \eqref{deqnc3}. This rules out case 3, and therefore the original ODE has no Liouvillian solutions.

\subsection*{NS5-P supertube NVE}
For the NS5-P supertube NVE, given by \eqref{STNVE},
the poles are order 2 and the order at infinity is 2, so once again, all three cases of the Kovacic algorithm must be checked. The poles are located at $\pm 1, \pm i\sqrt{n_5}$ and $\pm\sqrt{1-n_5+n_5^2 R^2/k^2}$ Note that the pole at $x=0$ is absent in the supertube NVE.

\noindent For case 1, we obtain
\begin{equation}
\begin{aligned}
       \{[\sqrt{U}]_c , \alpha_c^+ , \alpha_c^- \} &= \{ 0 , \ \frac34 , \ \frac14 \}  \\
        \{[\sqrt{U}]_\infty , \alpha_\infty^+ , \alpha_\infty^- \} &= \{ 0, \frac12, \frac12 \} \ ,
\end{aligned}
\end{equation}
for case 2, we obtain
\begin{equation}
\begin{aligned}
    E_c &= \{ 2,3,1 \} \\
    E_{\infty} &= \{ 2 \} \ ,
\end{aligned}
\end{equation}
and for case 3, we obtain
\begin{equation}
\begin{aligned}
    E_{c} &= \{3, 4,5,6,7,8,9 \} \\
    E_{\infty} &= \{ 6 \} \ .
\end{aligned}
\end{equation}
As in the case of the circular NS5 distribution, all three cases fail to produce a non-negative integer $d$ as defined in either \eqref{deqnc1},\eqref{deqnc2} or \eqref{deqnc3}. Thus the NVE has no Liouvillian solutions and string motion in the NS5-P supertube background is non-integrable.


\newpage

\bibliographystyle{JHEP.bst}

\bibliography{microstates}


\end{document}